\documentclass[11pt]{article}       
\usepackage{graphicx}
 \usepackage{mathptmx}      
\usepackage{amsmath} 
 
\def\overli{\overline} 
\def\overha{\widehat} 

\def\argmin \mathop{\rm argmin}
\def\Re{I\kern -0.37 em R}
\def\Na{I\kern -0.37 em N}
\def\Qe{I\kern -0.37 em Q}

\def\g{\,\vert \,}

\def\del{\partial\,}

\def\ev{\,\mbox{ev}\,} 
\def\evb{\overline{\ev}} 
\def\sev{\mbox{sev}} 
\def\sevb{\overline{\sev}} 
 
\def\zero{\mbox{\bf 0}}

\def\Chi2{\mbox{Q}} 
\def\pprod{\prod\nolimits}

\def\g{\,\vert\,}
\def\del{\partial\,}
\def\Pr{\mbox{Pr}}

\def\Tb{\overli{T}}
\def\Vb{\overli{V}}
\def\Wb{\overli{W}}
\def\sh{\overha{s}}
\def\th{\overha{\theta}}

\def\zero{\mbox{\bf 0}}

\def\wwhere{\mbox{ where }}

\def\Re{{R}}

\usepackage[T1]{fontenc} 
  
\usepackage{geometry}
\usepackage{fancyhdr}
\author{ {\large Carlos Alberto de Bragan\c{c}a Pereira}\vspace{2mm}\footnote{
ORCID: 0000-0001-6315-7484,  
ResearcherID: O-5022-2015.   
Institute of Mathematics of the Federal University of Matro Grosso do Sul. 
Av. Senador  Filinto M\"{u}ller, 1555, 79074-460, Campo Grande, Brazil. 
e-mail: \textit{cpereira@ime.usp.br} 			   
	}
    \\  
{\large Julio Michael Stern}\vspace{1mm}\footnote{ 
ORCID: 0000-0003-2720-3871,   
ResearcherID: C-1128-2013. 	   
Institute of Mathematics and Statistics 
of the University of S\~{a}o Paulo.  
Rua do Mat\~{a}o, 1010, 05508-900, S\~{a}o Paulo, Brazil.     
e-mail: \textit{jstern@ime.usp.br} \  
 (corresponding author)  
}
}

\title{The $e$-value\hspace{1pt}: \ A Fully Bayesian Significance Measure for  Precise Statistical Hypotheses and its Research Program}   	 

\begin{document}

\maketitle 

\maketitle

\centerline{Accepted for publication}  

\mbox{}

\begin{abstract}
This article gives a survey of the $e$-value, a  statistical significance measure a.k.a. the evidence rendered by observational data, $X$, in support of a statistical hypothesis, $H$, or, the other way around,  the epistemic value of $H$ given $X$. 
The $e$-value and the accompanying FBST, the Full Bayesian Significance Test, constitute the core of a research program that was started at IME-USP, is being developed by over 20 researchers worldwide, and has, so far, been referenced by over 200 publications.    
    
The $e$-value and the FBST comply with the best principles of Bayesian inference, including the likelihood principle, complete invariance, asymptotic consistency, etc. 
Furthermore, they exhibit powerful logic or algebraic properties in situations where one needs to compare or compose distinct hypotheses that can be formulated either in the same or in different statistical models. 
Moreover, they effortlessly accommodate the case of sharp or precise hypotheses, a situation where alternative methods often require ad hoc and convoluted procedures. 
Finally, the FBST has outstanding robustness and reliability characteristics, outperforming traditional tests of hypotheses in many practical applications of statistical modeling and operations research.   
         	
 
 
\end{abstract}

\section{Introduction}
\label{intro}

The Full Bayesian Significance Test (FBST) is a novel statistical test of hypothesis published in 1999 by both authors \cite{PereiraStern1999} and further extended in \cite{Madruga2003,Pereira2008}. 
This solution is anchored by a novel measure of \textit{statistical significance} known as the $e$-value, $\ev(H\g X)$, a.k.a. the \textit{evidence value} provided by observational data $X$ in support of the statistical hypothesis $H$ or, the other way around, the \textit{epistemic value} of hypothesis $H$ given the observational data $X$. 
The $e$-value, its theoretical properties and its applications have been a topic of research for the Bayesian Group at  USP, the University of S\~{a}o Paulo, for the last 20 years, including collaborators working at UNICAMP, the State University of Campinas, UFSCar, the Federal University of S\~{a}o Carlos, and other universities in Brazil and around the world. The bibliographic references list a selection of contributions to the FBST research program and its applications.  

The FBST was specially designed to provide a significance measure to \textit{sharp} or \textit{precise} statistical hypothesis, namely, 
hypotheses consisting of a zero-volume (or zero Lebesgue measure) subset of the parameter space.  
Furthermore the e-value has many necessary or desirable properties for a statistical support function, such as: 

(i) Give an intuitive and simple measure of significance for the 
hypothesis in test, ideally, a {\it probability} defined directly
in the  original or {\it natural parameter space}. 

(ii) Have an intrinsically geometric definition, independent of any
non-geometric aspect, like the particular parameterization of the
(manifold representing the) null hypothesis being tested, or the
particular coordinate system chosen for the parameter space, in short, be defined as an {\it invariant} procedure. 

(iii) Give a measure of significance that is smooth, i.e. 
{\it continuous and differentiable}, on the hypothesis parameters and 
sample statistics, under appropriate regularity conditions for the model.    

(iv) Obey the {\it likelihood principle} , i.e., the information gathered from observations should be represented by, and only by, the likelihood function, \cite{Berger1988,Pawitan2001,Wechsler2008}.

(v) Require {\it no ad hoc artifice} like assigning a positive prior
probability to zero measure sets, or setting an arbitrary initial
belief ratio between hypotheses. 

(vi) Be a {\it possibilistic} support function, where 
the support of a logical disjunction is the maximum support 
among the support of the disjuncts, see \cite{SternPereira2014}.  

(vii) Be able to provide a {\it consistent} 
test for a given sharp hypothesis. 

(viii) Be able to provide {\it compositionality} operations in  
complex models. 

(ix) Be an {\it exact} procedure, i.e., make no use of ``large sample'' 
asymptotic approximations when computing the $e$-value.

(x) Allow the incorporation of previous experience or expert's opinion 
via (subjective) {\it prior distributions}.

The objective of the next two sections is to recall  standard nomenclature and provide a short survey  of the FBST theoretical framework, summarizing the most important statistical properties of its  statistical significance measure, the $e$-value; these introductory sections follow closely the tutorial \cite[appendix A]{Stern2008b}, see also \cite{Pereira2008}.

\section{Bayesian Statistical Models} 
\label{BayesSM}

A standard model of (parametric) Bayesian statistics concerns an observed (vector) random variable, $x$, that has a \textit{sampling} distribution with a specified functional form, $p(x \g \theta)$, indexed by the (vector) parameter $\theta$.   
This same function, regarded as a function of the free variable $\theta$ with a fixed argument $x$, is the model's \textit{likelihood} function. 

In \textit{frequentist} or classical statistics, one is allowed to use probability calculus in the sample space, but strictly forbidden to do so in the parameter space, that is, $x$ is to be considered as a random variable, while $\theta$ is not to be regarded as random in any way. 
In frequentist statistics,  $\theta$ should be taken as a ``fixed but unknown quantity'', and neither probability nor any other belief calculus may be used to directly represent or handle the uncertain knowledge about the parameter.

In the Bayesian context, the parameter $\theta$ is regarded as a latent (non-observed) random variable. 
Hence, the same formalism  used to express 
(un)certainty or belief, namely, probability theory, is used in both the sample and the parameter space.   
Accordingly, the joint probability distribution, $p(x,\theta)$ should summarize all the information available in a statistical model.  
Following the rules of probability calculus, 
the model's joint distribution of $x$ and $\theta$ can be factorized either as the likelihood function of the parameter given the observation times the {\it prior} distribution on $\theta$, 
or as the {\it posterior} density of the parameter times the observation's marginal density,   
\[ 
p(x,\theta) = p(x \g \theta) p(\theta) 
= p(\theta \g x) p(x) \ . 
\]

The {\it prior} probability distribution $p_0(\theta)$  
represents the initial information available about the parameter.  
In this setting, a {\it predictive} distribution for 
the observed random variable, $x$, is represented by a mixture (or superposition) of stochastic processes, all of them with the functional form of the sampling distribution,according to the prior mixing (or weight) distribution,      
\[ 
p(x) = \int_{\theta} p(x \g \theta) p_0(\theta) d \theta \ . 
\]  

If we now observe a single event, $x$, it follows from the factorizations of the joint distribution above that the {\it posterior}  probability distribution of $\theta$, representing the available information about the parameter after the observation, is given by  
\[ 
p_1(\theta) \propto p(x \g \theta) p_0(\theta) \ . 
\]  

In order to replace the `proportional to' symbol, $\propto$, by an equality, it is necessary to divide the right hand side by the normalization constant,  
$c_1 = \int_\theta p(x \g \theta) p_0(\theta) d\theta$. 
This is the {\it Bayes rule}, giving the (inverse) probability of the parameter given the data. That is the basic learning mechanism of Bayesian statistics.  
Computing normalization constants is often difficult or 
cumbersome. 
Hence, especially in large models, it is customary to work with unormalized densities or {\it potentials} as long as possible in the intermediate calculations, computing only the final normalization constants.    
It is interesting to observe that the joint distribution function, taken with fixed $x$ and free argument $\theta$, is a potential for the posterior distribution.

Bayesian learning is a recursive process, where the 
posterior distribution after a learning step becomes the 
prior distribution for the next step. 
Assuming that the observations are i.i.d. (independent and identically distributed) the posterior distribution after $n$ observations, 
$x^{(1)},\ldots x^{(n)}$, becomes,             
\[ 
p_n(\theta) 
\ \propto \ 
p(x^{(n)} \g \theta) p_{n-1}(\theta) 
\ \propto \  \pprod_{i=i}^n 
p(x^{(i)} \g \theta) p_0(\theta) \ . 
\]  
If possible, it is very convenient to use a {\it conjugate prior}, that is, a mixing distribution whose functional form is invariant by the Bayes operation in the statistical model at hand.   
For example, the conjugate priors for the Normal and Multivariate models are, respectively, Wishart and the Dirichlet distributions, see \cite{Gelman2004,Zellner1971}.

The founding fathers of the Bayesian school, namely, Reverend Thomas Bayes,  Richard Price and Pierre-Simon de Laplace, interpreted the Bayesian operation as a path taken for learning about probabilities related to unobservable causes, represented by the parameters of a statistical model, from probabilities related to their consequences, represented by observed data.    
Nevertheless, later interpretations of statistical inference, like those of Bruno de Finetti who endorsed the epistemological perspectives of empirical positivism, strongly discouraged such causal interpretations, see \cite{Stern2017b,Stern2018a} for further discussion of this controversy.     

The `beginnings and the endings' of the Bayesian learning process deserve  further discussion, that is, we should present some rationale for choosing the prior distribution used to start the learning process, and some convergence theorems for the posterior 
as the number observations increases.    
In order to do so, we must access and measure the information content of a (posterior) distribution.  
\cite{Jeffreys1961,Kapur1989,Stern2011a,Zellner1971} explain how the concept of entropy can be used  to unlock many of the mysteries related to the problems at hand. 
In particular, they discuss some fine details about 
criteria for prior selection and important properties of posterior convergence.

\section{The Epistemic $e$-values} 
\label{eval}

Let $\theta \in \Theta \subseteq  \Re^p$ be a vector parameter 
of interest, and $p(x \g \theta)$ be the likelihood associated to 
the observed data $x$, as in the standard statistical model. 
Under the Bayesian paradigm the posterior density, 
$p_n(\theta)$, is proportional to the product of the 
likelihood and a prior density,  
\[ 
p_n(\theta) \propto p(x \g \theta) \, p_0(\theta) \ . 
\]  

A hypothesis $H$ states that the parameter lies in the 
null set, defined by inequality and equality constraints given by 
vector functions $g$ and $h$ in the parameter space,      
\[ 
\Theta_H = \{ \theta \in \Theta \g 
g(\theta) \leq \zero \wedge h(\theta) = \zero \} \ .
\] 
From now on, we use a relaxed notation, writing 
$H$ instead of $\Theta_H$.    
We are particularly interested in sharp (precise) hypotheses, i.e., 
those in which there is at least one equality constraint and, therefore, 
$\dim(H) < \dim(\Theta)$.

The FBST defines $\ev(H)$, the $e$-value supporting (in favor of) 
the hypothesis $H$, and $\evb(H)$, the $e$-value against $H$, as    
\[  
s(\theta) =   \frac{p_n(\theta)}{r(\theta)} \ , \ \   
s^* = s(\theta^*) = 
\sup\nolimits_{\theta \in H}  s(\theta) \ , \ \        
\sh = s(\th) = 
\sup\nolimits_{\theta \in \Theta}  s(\theta) \ ,         
\]   
\[  
T(v) =  \{ \theta \in \Theta \g s(\theta) \leq v \}   
\ , \ \  
W(v) = \int_{T(v)} p_n\left(\theta \right) d\theta 
\ , \ \ 
\ev(H) = W(s^*)  \ , 
\] 
\[  
\Tb(v) = \Theta - T(v) \ , \ \ 
\Wb(v) = 1-W(v)  \ , \ \  
\evb(H) = \Wb(s^*) = 1-\ev(H) \ .  
\] 

The function $s(\theta)$ is known as the posterior \textit{surprise function}  
relative to a given reference density, $r(\theta)$. 
$W(v)$ is the cumulative surprise distribution. 
Due to its interpretation in mathematical and philosophical logic, see \cite{Borges2007},  
$W(v)$ is also known as (the statistical model's)  
\textit{truth function} or \textit{Wahrheitsfunktion}.    
The surprise function was used in the context of  statistical inference by Good \cite{Good1983}, Evans \cite{Evans1997}, Royall \cite{Royall1997} and Schackle \cite{Shackle1968,Shackle1969}, among others. 
Its role in the FBST is to make $\ev(H)$ explicitly invariant under suitable transformations on the coordinate system of the parameter space, see next section. 

The tangential (to the hypothesis) set $\Tb=\Tb(s^*)$, is a Highest Relative Surprise Set (HRSS). 
It contains the points of the parameter space with higher surprise, relative to the  reference density, than any point in the null set $H$. 
When $r(\theta)\propto 1$, the possibly improper uniform density, $\Tb$ is the Posterior's Highest Density Probability Set (HDPS) tangential to the null set $H$. 
Small values of $\evb(H)$ indicate that the hypothesis traverses high density regions, favoring the hypothesis.

Notice that, in the FBST definition, there is an optimization step and an integration step. 
The optimization step follows a typical {\it maximum probability} argument, according to which, 
``a system is best represented by its highest probability realization''. 
The integration step extracts information from the system as a   
probability weighted average. 
Many inference procedures of classical statistics rely basically on 
maximization operations, while many inference procedures of Bayesian 
statistics rely on integration (or marginalization) operations. 
In order to achieve all its desired properies, the FBST procedure 
has to use both operation types.

\subsection{Nuisance Parameters} 
\label{nuisance}

Let us consider the situation where the hypothesis constraint, 
$H:\ h(\theta)=h(\delta)=0\ , \theta=[\delta,\lambda]$ is not a function
of some of the parameters, $\lambda$. 
This situation is described in \cite{Basu1988}  by Debabrata Basu as follows: 

\begin{quotation} 
	{\it ``If the inference problem at hand relates only to $\delta$, 
		and if information gained on $\lambda$ is of no direct relevance to the 
		problem, then we classify $\lambda$ as the Nuisance Parameter. 
		The big question in statistics is: How can we eliminate the 
		nuisance parameter from the argument?''}  
\end{quotation} 

Basu goes on listing at least 10 categories of procedures to  achieve
this goal, like using $max_\lambda$ or $\int \ d\lambda$,  the
maximization or integration  operators, in order to obtain a projected
profile or marginal  posterior function, $p(\delta \g x)$. 
The FBST does not follow the nuisance parameters elimination paradigm, 
working in the original parameter space, in its full dimension.

\subsection{Reference Prior and Invariance} 
\label{reference} 

In the FBST the role of the reference density, $r(\theta)$ is to make
$\evb(H)$ explicitly invariant under suitable transformations of
the coordinate system. 
The natural choice of reference density is an uninformative prior, interpreted as a representation of no information in the parameter space, or the limit prior for no observations, or the neutral ground state for the Bayesian learning operation.
Standard (possibly improper) uninformative priors include the uniform, maximum entropy densities, or Jeffreys' invariant prior. 
Finally, invariance, as used in statistics, is a metric concept, and the reference density can be interpreted as induced by the statistical model's information metric in the parameter space, $dl^2=d\theta'G(\theta)d\theta$, see \cite{Amari2007,Bernardo2005,Box1973,Fang1997,Gelman2004,Jeffreys1961,Kapur1989,Kapur1992,Zellner1971} for a detailed discussion. 
Jeffreys' invariant prior is proportional to the square root of the information matrix determinant,  
$p(\theta)\propto \sqrt{\mbox{det} G(\theta)}$.

\subsection*{Proof of invariance:} 

Consider a proper 
(bijective, integrable, and almost surely continuously 
differentiable) reparameterization 
$\omega=\phi(\theta)$. 
Under the reparameterization, the Jacobian, surprise, 
posterior and reference functions are:  
\[      
J(\omega)  =   
\left[ \frac{\del\theta}{\del\omega} \right] =  
\left[ \frac{\del \phi^{-1}(\omega)}{\del \omega} \right] =  
\left[ \begin{array}{ccc} 
\frac{\del\theta_1}{\del\omega_1} & \ldots & 
\frac{\del\theta_1}{\del\omega_n} \\ 
\vdots & \ddots & \vdots \\ 
\frac{\del\theta_n}{\del\omega_1} & \ldots & 
\frac{\del\theta_n}{\del\omega_n}  
\end{array} \right]    
\]  
\[ 
\widetilde{s}(\omega)   =    
\frac{ \widetilde{p}_n(\omega) }{ \widetilde{r}(\omega) } = 
\frac{ p_n(\phi^{-1}(\omega)) \left| J(\omega) \right| } 
{ r(\phi^{-1}(\omega))   \left| J(\omega) \right| } 
\] 

Let $\Omega_H = \phi(\Theta_H)$. 
It follows that 
\[ \widetilde{s}^* = 
\sup_{\omega \in \Omega_H} \widetilde{s}(\omega)  = 
\sup_{\theta \in \Theta_H}  s(\theta)  =  s^*  
\] 
hence, the tangential set, 
$\Tb \mapsto \phi(\Tb) = \widetilde{\Tb}$, and  
\[ 
\widetilde{\mbox{ev}}(H) = 
\int_{\widetilde{\Tb}} \widetilde{p}_n(\omega) d \omega = 
\int_{\Tb} p_n(\theta) d \theta = \evb(H) .  
\]

\subsection{Asymptotics and Consistency} 
\label{Asympt}

Let us consider the cumulative distribution 
of the evidence value against the hypothesis,  
$\Vb(c)= \Pr(\evb \leq c)$,   
given $\theta^0$, the true value of the parameter.  
Under appropriate regularity conditions, for increasing sample size, 
$n\rightarrow \infty$, we can say the following:

- If $H$ is false, $\theta^0\notin H$, then  
$\evb$ converges (in probability) to 1, that is,  
$\Vb(0\leq c <1)\rightarrow 0$.

- If $H$ is true, $\theta^0\in H$,  then $\Vb(c)$, 
the confidence level,  is approximated  by the function   
\[   
QQ(t,h,c) = 
\Chi2\left(t-h, \Chi2^{-1}\left(t,c\right) \right) \ , \ \ 
\wwhere 
\] 
\[ 
\Chi2(k,x) = 
\frac{\Gamma(k/2, x/2)}{\Gamma(k/2, \infty)} \ , \  \  
\Gamma(k,x) = \int_0^x y^{k-1}e^{-y} dy \ ,  
\] 
$t=\dim(\Theta)$, $h=\dim(H)$ and $\Chi2(k,x)$ is the 
cumulative chi-square distribution with $k$ degrees of freedom. 

Under the same regularity conditions, an appropriate choice of 
threshold or critical level, $c(n)$, provides a consistent test, 
$\tau_c\ $, that rejects the hypothesis if $\evb(H) > c$.    
The empirical power analysis developed in \cite{SternZacks2002,Lauretto2003} provides critical levels that are consistent and also effective for small samples.


\subsection*{Proof of consistency:}

Let  $\Vb(c)= \Pr(\evb \leq c)$  be the cumulative distribution of the
evidence value against the  hypothesis, given $\theta$. 
We stated that, under appropriate regularity conditions, for increasing
sample size, $n\rightarrow \infty$, if $H$ is true, i.e. $\theta\in H$,
then $\Vb(c)$, is approximated  by the function   
\[ 
QQ(t,h,c) = 
\Chi2\left(t-h, \Chi2^{-1}\left(t,c\right) \right) \ . 
\]

Let $\theta^0$, $\th$ and $\theta^*$ be the true value, 
the unconstrained MAP (Maximum A Posteriori), and constrained 
(to $H$) MAP estimators of the parameter $\theta$.  

Since the FBST is invariant, we can chose a coordinate system  where,
the (likelihood function) Fisher information matrix at the true
parameter value  is the identity, i.e., $J(\theta^0)=I$. 
From the posterior Normal approximation theorem, 
see \cite{Gelman2004},  we know that the standarized total
difference between $\th$ and $\theta^0$  converges in distribution  to 
a standard Normal distribution, i.e. 
\[ 
\sqrt{n}(\th -\theta^0) \rightarrow 
N\left( 0, J(\theta^0)^{-1} J(\theta^0) J(\theta^0)^{-1} \right) = 
N\left( 0, J(\theta^0)^{-1} \right) = 
N\left( 0, I \right) 
\] 

This standarized total difference can be decomposed into tangent
(to the  hypothesis manifold) and transversal orthogonal components, i.e. 
\[ 
d_t = d_h + d_{t-h} \ , \  
dt = \sqrt{n}(\th -\theta^0) \ , \    
d_h = \sqrt{n}(\theta^* -\theta^0) \ , \     
d_{t-h} = \sqrt{n}(\th -\theta^*)  \ .  
\] 
Hence, the total, tangent and transversal distances ($L^2$ norms), 
$||d_t||$, $||d_h||$ and $||d_{t-h}||$, 
converge in distribution to chi-square variates with, respectively,
$t$, $h$ and $t-h$ degrees of freedom.  

Also from, the MAP consistency, we know that the MAP 
estimate of the Fisher information matrix, $\widehat{J}$,  
converges in probability to true value, $J(\theta^0)$. 

Now, if $X_n$ converges in distribution to $X$, and $Y_n$ converges 
in probability to $Y$, we know that the pair $[X_n, Y_n]$ converges 
in distribution to $[X, Y]$. 
Hence, the pair $[||d_{t-h}||, \widehat{J}]$ converges in 
distribution to $[x, J(\theta^0)]$, where $x$ is a chi-square 
variate with $t-h$ degrees of freedom. 
So, from the continuous mapping theorem, the evidence value 
against $H$, $\evb(H)$, converges in distribution to 
$\overline{e}=\Chi2(t,x)$, where $x$ is a chi-square variate with 
$t-h$ degrees of freedom. 

Since the cumulative chi-square distribution is an increasing function, 
we can invert the last formula, i.e., 
$\overline{e}=\Chi2(t,x)\leq c \Leftrightarrow x \leq \Chi2^{-1}(t,c)$. 
But, since $x$ in a chi-square variate with 
$t-h$ degrees of freedom, 
\[ 
\Pr(\overline{e} \leq c) =  
QQ(t,h,c) = 
\mbox{Q.E.D.} 
\] 
A similar argument, using a non-central chi-square distribution, proves the other asymptotic statement. 

\subsection{Decisions:  Reject $H$, remain Neutral, or Accept} 
\label{Asympt}

In this subsection we briefly discuss the important question of deciding when to Accept, or Reject, or remain Neutral about a statistical hypothesis $H$, given observed data $X$. 
We start our discussion elaborating on the asymptotic results derived in the last sub-section. 

If a random variable, $x$, has a continuous cumulative distribution function, $F(x)$, its probability integral transform generates 
a uniformly distributed random variable, $u=F(x)$, see \cite{Angus1994}.  
Hence, the tranformation $\sevb =QQ(t,h,\evb)$, defines a ``standarized $e$-value'', $\sev =1-\sevb$, that can be used somewhat in the same way as a $p$-value of classical statistics. 
This standarized $e$-value may be a convenient value to report, since its asymptotically uniform distribution (under $H$) provides a large-sample limit interpretation, and many researchers will feel already familiar with consequent diagnostic procedures for scientific hypotheses based on adequately large empirical data-sets.   
In particular, a researcher may use cut-off thresholds already familiar to him when dealing with $p$-values.   
Efficient procedures for computing empirical cut-off thresholds that are effective for small size data sets are developed in    \cite{Bernardo2012,Lauretto2003,Lauretto2007,Lauretto2005a,Lauretto2005b}. 

Traditionally, statisticians are used to establish a dichotomy: Reject/ Accept (technically, Not-Reject) $H$ if the significance measure in use is below or above the established cut-off threshold. 
Nevertheless, a thorough analysis of consistent desiderata for logical properties of such a decision procedure take us to an unavoidable conclusion: The classical Reject/ Accept dichotomy must be replaced by a trichotomy, namely, Reject/ remain Neuter (a.k.a remain undecided or agnostic)/ Accept $H$ if, respectively,  
$0 \leq \sev(H\g X) < c_1$, 
$c_1 \leq \sev(H\g X) < c_2$, or 
$c_2 \leq \sev(H\g X) \leq 1$, where 
$0 < c_1 < c_2 < 1$; 
For an extensive and detailed analysis of consistent desiderata for statistical test procedures, see \cite{Izbicki2015,Silva2015}. 

The study of such logical desiderata was in part motivated as a way to contrast the statistical properties of the FBST with other statistical tests of hypotheses.   
Surprisingly, it is possible to travel this path in the opposite direction, that is, it is possible to start from consistent desiderata for logical properties of statistical tests and, from those, derive a complete characterization of a class of statistical significance measures and hypothesis tests that coherently generalizes the FBST, see \cite{Esteves2016,Esteves2019,Stern2018c} for further details. 
Moreover, this Generalized FBST finds interesting applications in metrology and related fields, were reliable bounds for the precision of experimental measurements can be obtained from sources external to the statistical experiment designed to test the hypothesis under scrutiny. 
Finally, this kind of detailed error analysis for crucial scientific experiments finds valuable applications in the fields of metrology, epistemology, and philosophy of science, see \cite{Esteves2019} and future research.

\section{A Survey of FBST Related Literature}
\label{Literature}

A systematic cataloging of all published articles related to this research program is beyond the scope of this article; in the next subsections we survey a selection of such articles. 
The selected articles provides a sample covering diverse areas like statistical theory and methods, applications  to statistical modeling and operations research, and research in foundations of statistics, logic and epistemology.  
This selection is certainly biased, favoring the the authors' personal taste or involvement.

\subsection{Statistical Theory}
\label{Theory}

Several authors have developed the statistical theory that provides the mathematical formalism and demonstrates the outstanding statistical properties  FBST and its significance measure, the $e$-value.   
The following articles have explored and developed these themes of research:

\begin{itemize} 
	
	\item \cite{PereiraStern1999} is the first article of this research program. It presents the basic definition of the $e$-value and the FBST, and gives several simple and intuitive applications. 
	\cite{Madruga2003} provides an explicitly invariant version of the inference procedures. 
	After a long process in which the authors had to overcome objections raised by influential mainstream Bayesian thinkers, \cite{Pereira2008} was published in the flagship journal of ISBA - the \textit{International Society for Bayesian Analysis}.  
	\cite{Pereira2011} provides an entry on the FBST in the 
	\textit{International Encyclopedia of Statistical Science}. 
	
	\item \cite{Assane2018,Assane2019,Lauretto2007,Lauretto2005a,Lauretto2005b,Pereira2b} give and extensive treatment for the case of nonnested and separate hypotheses, including a detailed analysis  some of Bayesian classifiers.   
	
	\item \cite{DCunha2016,Diniz2012b,Loschi2012,Patriota2013,Patriota2017,Silva2018} establish several theoretical or empirical relations between the the $e$-value and alternative significance measures.       
	
	\item \cite{Cabras2015,Pinto2012,Ranzato2018,Ruli2020,Ruli2020b,Ventura2013,Ventura2014,Ventura2016} develop higher order asymptotic approximations of (log) likelihood and pseudo-likelihood functions that, in turn, are used do develop high-precision but fast computational algorithms for calculating $e$-values in parametric models.  
	The availability of a good library of such fast and reliable computer programs will, in turn, we believe, facilitate the incorporation of the FBST in statistical softwares intended for end-users or routine applications.

\end{itemize} 

\subsection{Statistical Modeling}
\label{Modeling}

Several authors have developed a wide range of applications of the FBST to statistical modeling and operations research.  
The following articles have explored and developed these themes of research:

\begin{itemize}  
	
	\item \cite{PereiraStern1999b} applies the FBST to software compliance testing and certification.  
	
	\item \cite{Lauretto2003} provides a unified and coherent treatment to a large class of  structural models based on the multivariate normal distribution. Previously, sub-classes of these models had to be handled individually using tailor-made tests. \cite{Vikas2016} gives some simple applications. 
	
	\item \cite{Diniz2011,Diniz2012a,Chen2015,Vosseler2016} develop or use unit root and cointegration testing for time series. The FBST is shown to be more reliable and effective than several previously published tests, without the need of any ad hoc artifices, like specially designed artificial priors (an obvious oxymoron).
	
	\item \cite{Irony2002,Maranhao2012,Rodrigues2006} apply the FBST to failure analysis and systems' reliability.  
	
	\item \cite{Cerezetti2012} applies the FBST to detect equilibrium conditions, or the lack thereof, in market prices of economic commodities or financial derivative contracts. 
	
	\item \cite{Chaiboonsri2018} applies the FBST in the context of empirical economic studies.   
	
	\item \cite{Garcia2016} use the FBST for selection and testing of statistical copulas. 
	
	\item \cite{Hubert2009,SternZacks2002} consider applications using generalizations of the Poisson distribution. 
	
	\item \cite{Hubert2018,Hubert2018b,Hubert2019} use the FBST for signal processing and detection of acoustic events.  
	
	\item \cite{Sikov2019} applies the FBST to model selection in statistical studies conducted 
	under informative sampling conditions. 
	
	\item \cite{Izbicki2012,Lauretto2009,Loschi2007,Montoya2001,Nakano2006,Rincon2010,Wittenburg2016} use the FBST to  
	verify Hardy-Weinberg equilibrium conditions, and other applications of statistical modeling in the area of genetics. 
	
	\item \cite{Andrade2015,Bernardini2011,Kostrzewski2012,Rifo2009,Rifo2012} use the FBST to test parametric hypotheses related to generalized Brownian motions, continuous or jump diffusions, extremal distributions, persistent memory and other stochastic processes.   
	
	\item \cite{Andrade2014,Bernardo2012,Oliveira2018,PereiraStern2008b} develop theory or applications of the FBST for statistical hypotheses related to independence in contingency tables and other multinomial models.
	
	\item \cite{Ainsbury2013,Camargo2012,Kelter2020,Lima2014,Mathis2008,Pereira2a,PereiraStern2001,Santos2020,Seixas2008,Spektor2019} apply the FBST for checking hypotheses in statistical models applied to biological sciences, ecology, environmental sciences,  medical diagnostics and efficacy evaluation, psychology and psychiatry.   
	
	\item \cite{Chakrabarty2017,Johnson2009} apply the FBST to test hypotheses in astronomy and astrophysics. 
	
\end{itemize}

\subsection{Foundations of Statistics, Logic and Epistemology}
\label{Foundations} 

Traditional significance measures used in statistics are always designed to work in tandem with a specific epistemological framework that gives them an appropriate interpretive context and support.  
For example, $p$-values are usually presented in the context of the ``judgment metaphor'' and the deductive falibilism epistemological framework, as developed by the philosopher Karl Popper, among others. 
Meanwhile, Bayes factors are presented in the context of the ``gambling metaphor'' and utility based decision theory, as developed by Bruno de Finetti, see \cite{Dubins1965,Finetti1975,Kadane2011,Kadane2016}. 
Furthermore, the logic or algebraic properties of each significance measure, in its appropriate domain of statistical hypotheses, must be mutually supportive and compatible with intended interpretations. 
The following articles have explored and developed these themes of research:

\begin{itemize}  
	
	\item \cite{Madruga2001,Silva2015,Thulin2014} analyze the FBST from a decision-theoretic Bayesian perspective. The first paper proves the ``Bayesianity'' of the FBST, in the sense its inference procedures can be derived by minimization of an appropriate loss function.   
	
	\item \cite{Madruga2003,SternPereira2014} compare the theoretical properties of the $e$-value with those of traditional significance measures, like the $p$-value and Bayes Factors.
	These articles analyze in great detail historical arguments given by celebrated statisticians against the use of procedures based on highest density probability sets. 
	Among those that opposed such ideas is Dennis Lindley, an influential figure at IME-USP and a personal friend of the first author. 
	Finally, \cite{Madruga2003,SternPereira2014} analyze  historical desiderata for an acceptable Bayesian significance test that were formulated by the frequentist statistician Oscar Kempthorne to the first author, and show how the FBST successfully achieves all these desired characteristics.   
	
	\item  \cite{Borges2007} analyzes the composition of hypotheses defined in independent statistical models and the corresponding composition rules for $e$-values and truth functions. 
	
	\item 
	\cite{Vieland2019} studies significance measures for evidence amalgamation and meta-analysis.  
	
	\item \cite{Esteves2019,Fossaluza2017,Izbicki2015,Esteves2016,Stern2018c} analyze conditions of logical consistency for significance measures and test procedures for several hypotheses defined in the same statistical model. 
	Conversely, these articles fully characterize some (agnostic or trivalent) generalizations of the FBST as the only statistical tests satisfying such logical consistency conditions.    
	
	\item  \cite{Stern2007a,Stern2007b,Stern2008a,Stern2011a,Stern2011b,Stern2014,Stern2015,Stern2017a} develop the Objective Cognitive Constructivism as an epistemological framework formally compatible and semantically amenable to the $e$-value significance measure and the FBST hypothesis test. 
	
	\item 
	\cite{Cristofaro2003} analyzes solutions to the problem of (statistical) induction, including Bayesian perspectives in general and the FBST in particular. 
	
	\item \cite{Hubert2018b,Pigliucci,Stern2003,Stern2004,Stern2018a} apply  concepts related to the FBST or the Objective Cognitive Constructivism epistemological framework to the study of economic or legal systems.  
	
	\item \cite{Stern2017b,Stern2018b} analyze the philosophical premises used by Karl Pearson to  
	define the $p$-value and to establish the epistemological foundations of frequentist statistics;  why Pearson's work and the subsequent work of Bruno de Finetti reversed previous commitments of Bayesian statistics; and how the FBST can be seen a way to reenter the path envisioned by the founding fathers of the Bayesian school, namely, Reverend Thomas Bayes,  Richard Price and Pierre-Simon de Laplace.  
	
	\item \cite{Bonassi2009,Fossaluza2015,Lauretto2012,Marcondes2019,Saa2019,Stern2008a} analyze the role of randomization procedures in the context of the Objective Cognitive Constructivism epistemological framework in particular, and in Bayesian statistics in general. 
	
\end{itemize}

\section{Future Research and Final Remarks}

The FBST research program has grown and spread far and wide, in some directions suggested by these authors, and also in other directions that were for us completely unforeseen and wonderfully surprising.   
We are confident that this research program will continue to flourish and expand, exploring new areas of theory and application. 
The authors would like to suggest a few topics (focusing on theoretical and applied statistics) worthy of further attention as possible entry points for those interested (be all welcome) in joining this research program:

\begin{itemize}  
	
	\item (1) In the context of information based medicine, see \cite{Pereira2a}, it is important to compare and test the sensibility and specificity of alternative diagnostic tools, access the bio-equivalence of drugs coming from different suppliers, identify and test the efficacy of possible genetic markers for clinical conditions, etc.  
	How to combine fast and computationally inexpensive heuristic algorithms and reliable statistical test procedures to best handle these and similar problems?       
	
	\item (2a) Influence diagrams are a powerful tool for decision modeling, see \cite{Barlow1993,Pereira1990}. Nevertheless, it is often hard to select optimal diagrams to model complex applications, see for example \cite{Mathis2008,Shavitt2017}. 
	How can the FBST best be used for sequential or concomitant inclusion/ exclusion of links or edge selection in influence graphs?     
	
	\item (2b) The aforementioned questions also arise in the context of Bayesian networks. In this context, it is important not only to test the significance of individual edges, but also to test the integrity of higher level sparsity structures, like the network click structure or its block factors, see  \cite{Stern1992,Stern2008a,SternColla2009,SternVavasis1994}. 
	
	\item (3) The $e$-value and the FBST were originally developed for parametric models. How can the $e$-value be used, interpreted, computed (and maybe generalized) in semi-parametric or non-parametric settings?     
	For instance, in models using functional bases, how can we test speeds of convergence for series expansions?   
	
	\item 
	(4a) The compositionality rules established in \cite{Borges2007} are based on functional operations over the truth functions, $W(v)$. \cite{Barlow1981,Kaufmann1977} present similar rules (for serial-parallel composition) in the context of reliability theory. Can these theories be seen as particular cases of more general and abstract logical formalisms?     
	
	\item 
	(4b) The same compositionality rules assume independence between distinct models in a given structure. Could statistical copulas, see \cite{Fossaluza2018,Garcia2016,Kaplan1987,Nelsen2006}, be used to successfully capture weak dependencies between distinct truth functions?   
	
	\item (5) The conditions for pragmatic acceptance of sharp hypotheses stated in \cite{Esteves2019} depend on consensual bounds for background uncertainties. 
	For universal physical  constants, metrologists establish such bounds by aggregating results of diverse experiments; similar situations occur in meta-analysis studies. 
	Several statistical methods have been proposed to aggregate such diverse data-sets, see \cite{Cohen1957,Garcia2003,Kelley1987a,Kelley1987b,Minka2005}. 
	What are the best ways to coherently establish and represent aggregate uncertainty bounds in the FBST framework?

\end{itemize}

\textbf{Acknowledgments:} 
The authors are grateful to IME-USP - the Institute of Mathematics and Statistics of the University of S\~{a}o Paulo, and INMA-UFMS - the  Institute of Mathematics of the Federal University of Mato Grosso do Sul. 
The authors are extremely grateful for the support received from their colleagues, collaborators, users and critics in the construction works of this research project.   

\textbf{Funding:} This research was funded by 
CNPq - the Brazilian National Counsel of Technological and Scientific Development (grants PQ 302767/2017-7, PQ 301892/ 2015-6); and  
FAPESP - the State of S\~{a}o Paulo Research Foundation (grants CEPID Shell-RCGI 2014/ 50279-4, CEPID CeMEAI  2013/07375-0). 

\textbf{Conflicts of Interest:} The authors declare no conflict of interest. The funders had no role in this study's data analyses, in its methodological developments or conclusions, in writing this manuscript, or in deciding where to publish. 

\textbf{Author Contributions:} All authors contributed equally to this paper.

 \renewcommand{\baselinestretch}{0.99}
\parskip 0.76mm
\begin{small} 


 \end{small} 
 
\end{document}